\documentstyle[aps,prd,epsfig,preprint,tighten]{revtex}
\begin{document}
\draft
\title{	Day-night asymmetry \\ 
	of high and low energy solar neutrino events\\ 
	in Super-Kamiokande and in the Sudbury Neutrino Observatory}
\author{ G.L.\ Fogli~$^a$, E.\ Lisi~$^a$, D.\ Montanino~$^b$, 
and A.\ Palazzo~$^a$}
\address{$^a$~Dipartimento di Fisica and Sezione INFN di Bari\\
             Via Amendola 173, I-70126 Bari, Italy \\ }
\address{$^b$~Dipartimento di Scienze dei Materiali dell'Universit\`a di Lecce\\
             Via Arnesano, I-73100 Lecce, Italy \\ }
\date{\today}
\maketitle
\begin{abstract}%............................................................
In the context of solar $\nu$ oscillations among active states, we briefly
discuss the current likelihood of Mikheyev-Smirnov-Wolfenstein (MSW)  solutions
to the solar $\nu$ problem, which appear to be currently favored at large
mixing, where small  Earth regeneration effects might still be observable in
Super-Kamiokande (SK) and in the Sudbury Neutrino Observatory (SNO). We point
out that, since such effects are larger at high (low) solar $\nu$ energies for
high (low) values of the mass square difference $\delta m^2$, it may be useful
to split the night-day rate asymmetry  in two separate energy ranges. We show
that the difference  $\Delta$ of the night-day asymmetry at high and low energy
may help to discriminate the two large-mixing solutions at low and high 
$\delta m^2$ through a sign test, both in SK and in SNO, provided that the
sensitivity to  $\Delta$ can reach the (sub)percent level.  
\end{abstract}%.............................................................
\medskip\medskip

%%%%%%%%%%%%%%%%%%%%%%%%%%%%%%%%%%%%%%%%%%%%%%%%%%%%%%%%%%%%%%%%%%%%%%%%%%%%%%
\section{Current MSW solutions: LMA, LOW, SMA}

The   well-known Mikheyev-Smirnov-Wolfenstein (MSW) \cite{Wo78} solutions to
the solar $\nu$ problem \cite{Ba89}  appear to have a somewhat different
likelihood after the most recent data from the Super-Kamiokande (SK)
experiment  \cite{Su00}, as compared with previous fits \cite{Fo00,Ba00,Go00}. 
The new SK data, both by themselves \cite{Su00}  and in combination with the
results of the chlorine (Cl) \cite{La00} and gallium (Ga) experiments  (SAGE
\cite{Ga00}, GALLEX  and GNO \cite{Be00}), tend now to favor the so-called
large mixing angle (LMA) and low mass (LOW) regions of the parameter space,
with respect to the region at small mixing angle (SMA) \cite{Su00}. This
tendency emerges mainly from a ``tension'' between the flat  (normalized)
spectrum observed in SK and the prediction of a nonvanishing spectral
distortion in the SMA region favored by total rates \cite{Su00}. However, a
compromise between such two discordant indications is still reached in global
fits.

Figure~1 shows the results our global $2\nu$ (active) oscillation fit to the
latest 39 solar neutrino data, including the three SK, Cl, and Ga total rates
\cite{Su00,La00,Ga00,Be00},  and the two 18-bin day and night SK energy spectra
of events \cite{Su00,SuPe} above 5.5 MeV. We take as  free variables  the
overall spectrum  normalization   (to avoid double counting of the total SK
rate information), the $\nu$ mass square difference $\delta m^2$, and the
mixing angle $\omega$, parametrized in terms of $\tan^2\omega$ to cover the
full range $\omega\in [0,\pi/4]$  \cite{Fo96}. Details of our calculations and
of the $\chi^2$ statistical analysis  can be found in \cite{Fo3n} and
references therein. Subleading quasi-vacuum effects at low $\delta m^2$
\cite{Pe88,Pa90}, recently revisited in  \cite{Fr00}, are taken into account as
described in \cite{FoQV}.

The three local $\chi^2$ minima turn out to be $\chi^2_{\min}=35.1$ (LMA), 38.7
(LOW), and 40.7 (SMA). This implies, from the point of view of {\em hypotheses
tests\/} \cite{PDG0} ($N_{\rm DF}=39-3$), that any of  the three solutions is
acceptable, since $\chi^2_{\min}/N_{\rm DF}\sim 1$ in any case. From the point
of view of mass-mixing neutrino {\em parameter estimation\/}  \cite{PDG0}
($N_{\rm DF}=2$), the  relative likelihood is instead governed by
$\Delta\chi^2=\chi^2-35.1$. In Fig.~1, $\Delta\chi^2$ contours at 90\%, 95\%,
and 99\% C.L.\ are shown as thin solid, thick solid, and dotted curves,
respectively. The LOW solution emerges at 84\% C.L.\ ($\Delta\chi^2=3.6$),
while the SMA solution emerges only at 94\% C.L. ($\Delta\chi^2=5.1$).

Figure~2 shows, for the sake of completeness, the regions favored by fits to
total rates (Cl+Ga+SK), and by the SK day-night spectra, at the same C.L.'s as
in Fig.~1. As already mentioned \cite{Su00}, a ``tension'' emerges between the
total rate data (which are highly consistent with the SMA solution) and the SK
spectral data (which disfavor such solution).%
%-------------------------------------------------------------------------
\footnote{ Notice that, conversely, the LOW solution is disfavored by total
rates, but is highly consistent with SK day-night spectra.}
%-------------------------------------------------------------------------
However, this tension is not enough to exclude the SMA region yet. In fact, a
compromise is reached in the global fit of Fig.~1 where, as compared with
Fig.~2, the SMA solution survives at smaller mixing angles, corresponding to
smaller spectral distortions. The price to pay is an  increase in the C.L.\ at
which the SMA solution emerges  (94\% in Fig.~1),  as compared with previous
analyses \cite{Fo00,Ba00,Go00}.%

Summarizing, the LMA and LOW solutions appear to be favored over the SMA
solution in the global fit,  although it is rather premature to think that the
latter is ruled out.  In the following, we discuss a possible way to
discriminate the two most likely solutions (LMA and LOW) in favorable
situations, by separating Earth regeneration effects in two distinct energy
ranges.

%%%%%%%%%%%%%%%%%%%%%%%%%%%%%%%%%%%%%%%%%%%%%%%%%%%%%%%%%%%%%%%%%%%%%%%%%%%%%%
\section{A possible test to discriminate the LOW and LMA solutions}

The slightly positive indication $(\sim 1.3\sigma)$  for an excess of nighttime
to daytime events in Super-Kamiokande \cite{Su00}, if confirmed with higher
statistical significance,  would indicate the occurrence of the Earth
regeneration effect  for $^8$B solar neutrinos (see, e.g., 
\cite{Li97,Ma00,Kr97,Gu99,Sm00}  for recent night-day asymmetry studies in SK).
Such indication, by itself, might not be sufficient to discriminate the LMA
from the LOW solution, since a slight excess is predicted in both cases.
However, it has long been known that the Earth regeneration effect for solar
neutrinos is strongly dependent on  the neutrino energy \cite{Ba80,Mi86}. Such
dependence  leads to several effects that might be observed in the
Super-Kamiokande experiment \cite{Su00} and in the Sudbury Neutrino Observatory
(SNO) \cite{Mc00}, including night-day variations of energy spectrum
distortions \cite{Li97,Gu99,Kr00},  or variations of the night-day rate
asymmetry with the electron energy threshold \cite{Ma00,Kr97}. In particular,
starting from the simple observation that the Earth regeneration effect is
stronger at low energy for the LOW solution, and at high energy for the LMA
solution,%
%----------------------------------------------------------------------------
\footnote{See, e.g., Fig.~4 of \protect\cite{Kr97}.}
%---------------------------------------------------------------------------
we point out that it may be useful to study the night-day asymmetry in two
separate energy ranges in both SK and SNO.

For definiteness, we consider the two following representative ranges for the
total (measured) energy of recoiling electrons in SK ($\nu$-$e$ scattering) and
in SNO ($\nu$-$d$ absorption),
%..........................................................................
\begin{eqnarray}
{\rm Low\ range\ }(L) &=& [5,7.5]{\rm\ MeV}\ ,\\
{\rm High\ range\ }(H) &=& [7.5,20]{\rm\ MeV}\ ,
\end{eqnarray}
%..........................................................................
and  calculate the night-day rate asymmetry in such ranges,%
%----------------------------------------------------------------------------
\footnote{The 5 MeV threshold has already been reached in SK, although 
the $[5,5.5]$ MeV bin is not used yet in SK fits \cite{Su00}.}
%------------------------------------------------------------------------------
%.............................................................................
\begin{equation}
A_{L,H} = \left( \frac{N-D}{N+D}\right)_{L,H}\ .
\end{equation}
%.............................................................................
Since one expects $A_H\gtrsim A_L$ for the LMA solution and $A_H\lesssim A_L$
for the LOW solution, it is useful to introduce the difference
%.............................................................................
\begin{equation}
\Delta =  \left( \frac{N-D}{N+D}\right)_{H}-\left( \frac{N-D}{N+D}\right)_{L}\ ,
\end{equation}
%.............................................................................
which should change sign when passing from the LMA region
$(\Delta \gtrsim 0)$ to the LOW region $(\Delta \lesssim 0)$.

Figures~3 and 4 show the results of our calculations of $\Delta$  (eccentricity
effects removed) in SK and SNO,  respectively, in the form of isolines at 
$\Delta\times 100=\pm0.5$, $\pm1$, and $\pm 2$. Notice that the magnitude of
$\Delta$ in SNO is typically a factor of two higher than in SK. Figures~3 and 4
confirm that $\Delta>0$ ($<0$) would represent clear evidence in favor of the
LMA (LOW) solution, both in SK and SNO, thus allowing a useful ``sign
discrimination test'' to solve the LOW-LMA ambiguity at large mixing. The power
of such test decreases as  $\Delta\to 0$  in the upper part (lower part) of the
LMA (LOW) solution, corresponding to vanishing Earth regeneration effects for
$^8$B neutrinos in SK and SNO.

From a comparison of Figs.~1, 3, and 4, it turns out that, in the most
favorable case for the LOW solution (i.e., in its upper part),  the quantity
$\Delta\times 100$ can approximately reach the value $-0.5$ in SK and $-1$ in
SNO; analogously, it can reach the value $+1$ (SK) and $+2$ (SNO) for the LMA
solution. The separation  between the two solutions is thus $\Delta({\rm
LMA})-\Delta({\rm LOW}) \lesssim 1.5\%$ in SK ($\lesssim3\%$ in SNO) and, in
its upper range, it seems not  too far from the present  experimental
sensitivity to day-night effects.%
%------------------------------
\footnote{The quoted SK uncertainty on $A=N-D/N+D$, integrated over  the full
energy range, is $\pm 1.1\%$~(stat.)~$\pm0.6\%$~(syst.)  \protect\cite{Su00}.
The (larger) total uncertainty of $A_H$ and $A_L$ in the two  $H$ and $L$
energy sub-ranges is presumably at the $\sim 2\%$ level, although we have not
enough information for a precise estimate.}
%----------------------------- 

In general, however, one should require for detection of $\Delta\neq 0$ a
typical sensitivity  at the subpercent level in SK, and at the percent level in
SNO, whose viability requires not only a very high statistics, but also a
dedicated study of systematics (and of their cancellations in a difference like
$\Delta$). Notice also that the energy value separating the $L$ and $H$ ranges
does not need to be equal to our representative choice (7.5 MeV) nor to be the
same in SK and SNO, and should be tuned to optimize the  statistical
significance of the $\Delta$-sign test. The test would in any case benefit from
a reduction of the SK and SNO energy thresholds, which would both increase the
statistics and enhance the sensitivity to Earth effects at low energies.%
%-----------------------------------------------------------------------------
\footnote{Experiments sensitive to neutrino energies below the SK or SNO 
threshold can also observe relevant day-night effects in the LOW region
\protect\cite{Ra91} (see also \protect\cite{Fo00,Kr97} and references therein).}
%-----------------------------------------------------------------------------
Such
detailed experimental studies are  beyond the scope of this work, whose
main purpose is to emphasize that the difference of the night-day asymmetry at
low and high neutrino energy in SK and SNO might be observable in  favorable
situations, and help to separate the LOW and LMA solutions. Even in the absence
of an accurate measurement of $\Delta$, the simple {\em concordance\/}  of the
$\Delta$ sign in SK {\em and\/} SNO  ($++$ for the LMA solution or $--$ for
the LOW solution) would be a valuable information.

%%%%%%%%%%%%%%%%%%%%%%%%%%%%%%%%%%%%%%%%%%%%%%%%%%%%%%%%%%%%%%%%%%%%%%%%%%
\section{Summary and conclusions}

We have presented a global MSW oscillation fit to the solar neutrino data,
showing that all the three usual solutions  (SMA, LMA, and LOW) emerge at 95\%
C.L.\   The LOW and LMA solutions appear to be globally favored,  the latter
providing the best fit. Within the LMA (LOW) solution, the excess of nighttime
events due to $\nu_e$ regeneration in the Earth is larger in the higher (lower)
part of the neutrino energy range. Therefore, by taking the difference $\Delta$
of the night-day asymmetry in two separate ranges at high and low energy, a
small positive (negative) signal would provide evidence for the LMA (LOW)
solution.  Detection  of $\Delta\neq 0$ requires a (sub)percent sensitivity to 
day-night effects, which needs to be assessed by further experimental
investigations. However, the simple preference (and concordance) of the SK and
SNO experiments for a definite $\Delta$ sign would already provide us with
valuable information,  corroborating other tests envisaged to solve the LOW-LMA
ambiguity in such experiments \cite{Kr00}.

%%%%%%%%%%%%%%%%%%%%%%%%%%%%%%%%%%%%%%%%%%%%%%%%%%%%%%%%%%%%%%%%%%%%%%%%%%%%%%%
% 			R E F E R E N C E S 
%%%%%%%%%%%%%%%%%%%%%%%%%%%%%%%%%%%%%%%%%%%%%%%%%%%%%%%%%%%%%%%%%%%%%%%%%%%%%%%

%
%%%%%%%%%%%%%%%%%%%%%%%%%%%%%%%%%%%%%%%%%%%%%%%%%%%%%%%%%%%%%%%%%%%%%%%%%%%%%%%
%%%%%%%          P O S T S C R I P T       F I G U R E S 
%%%%%%%   memo:  to include them add epsfig in the \documentstyle
%%%%%%%          and move this part befor \end{document}. 
%%%%%%%          Include the following \newcommand:
%%----------------------------------------------------------------------------
\newcommand{\InsertFigure}[2]{\newpage\vspace*{-2.56cm}\begin{center}\mbox{%
\epsfig{bbllx=1.4truecm,bblly=1.3truecm,bburx=19.5truecm,bbury=26.5truecm,%
height=21.4truecm,figure=#1}}\end{center}\vspace*{-1.8truecm}%
\parbox[t]{\hsize}{\small\baselineskip=0.5truecm\hspace*{0.5truecm} #2}}
%----------------------------------------------------------------------------
\newcommand{\InsertFigures}[2]{\newpage\begin{center}\mbox{%
\epsfig{bbllx=1.4truecm,bblly=1.3truecm,bburx=19.5truecm,bbury=26.5truecm,%
height=21.4truecm,figure=#1}}\end{center}\vspace*{-3.8truecm}%
\parbox[t]{\hsize}{\small\baselineskip=0.5truecm\hspace*{0.5truecm} #2}}
%----------------------------------------------------------------------------
%%%%%%%%%%%%%%%%%%%%%%%%%%%%%%%%%%%%%%%%%%%%%%%%%%%%%%%%%%%%%%%%%%%%%%%%%%%%%%%
%..............................................................................
\InsertFigure{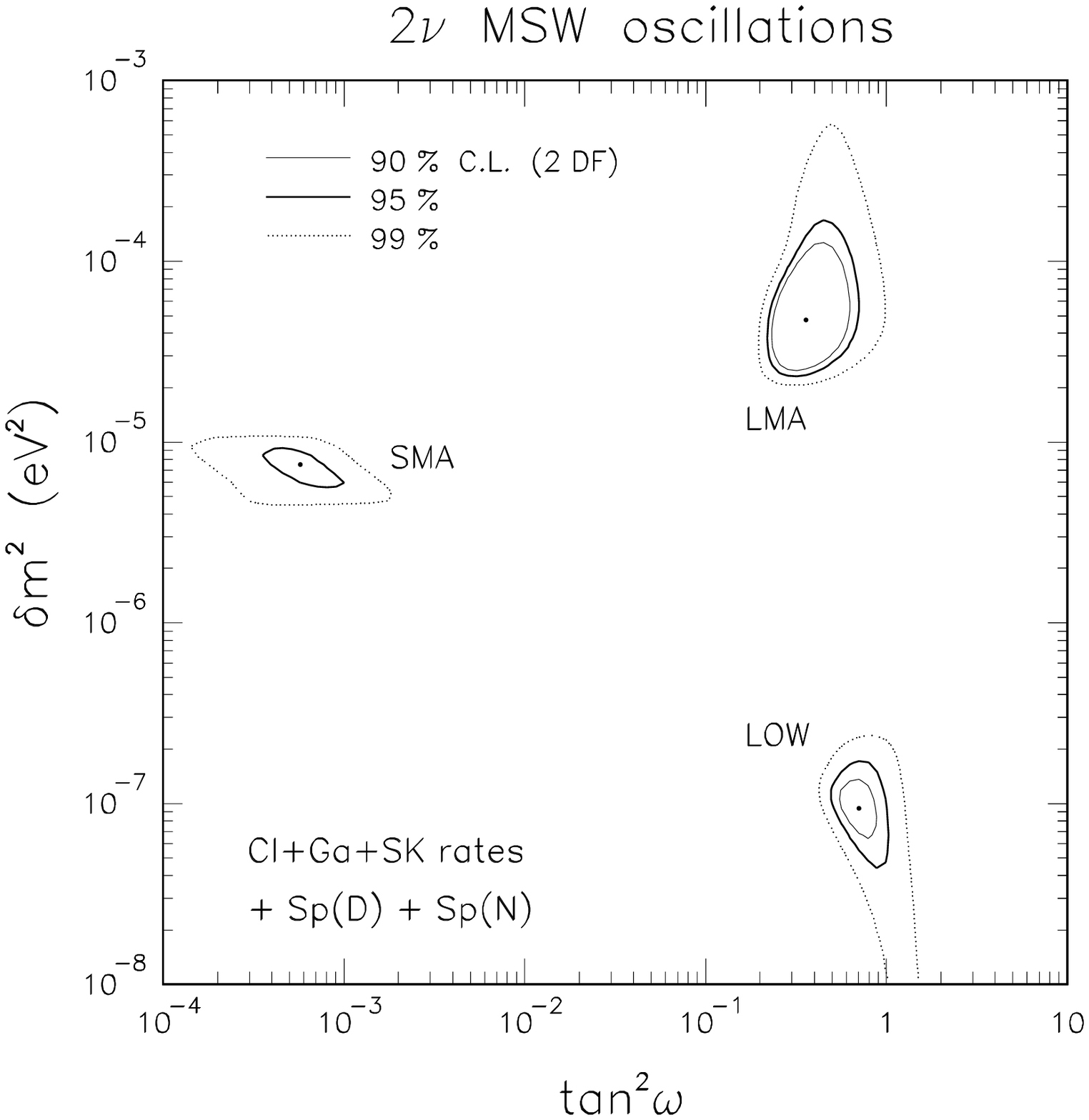}%
{Fig.~1. Global $2\nu$ fit including total solar neutrino  rates (Ga+Cl+SK)
\protect\cite{Su00,La00,Ga00,Be00} and the SK day and night spectra
\protect\cite{Su00,SuPe}.}
%%..............................................................................
\InsertFigure{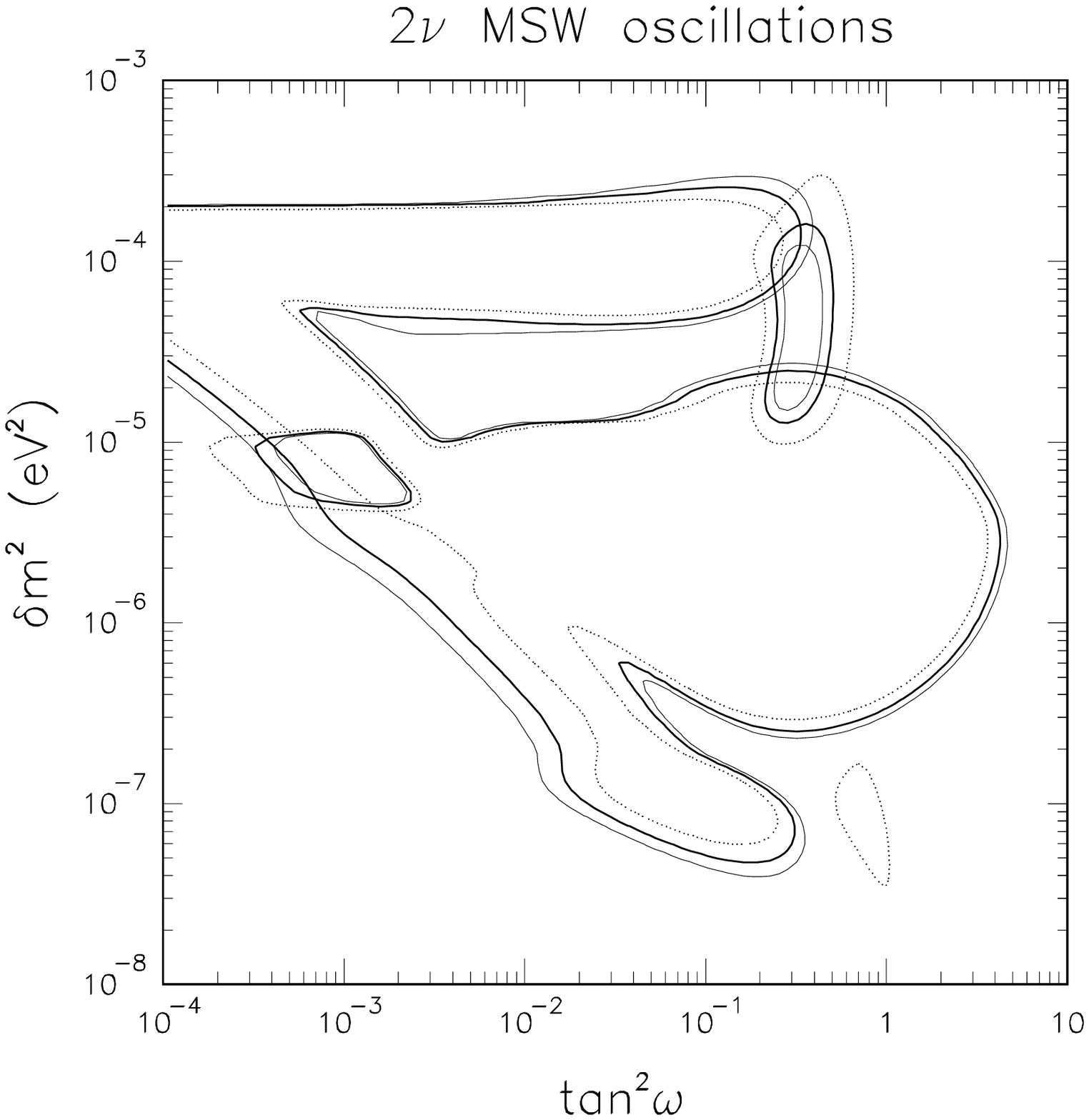}%
{Fig.~2. Results of the separate fits to total rates and to SK day-night
spectra. Total rate data mainly  determine the three SMA, LMA, and LOW allowed
regions, while the SK day-night spectra exclude a large part of the parameter
space where either spectral distortions or day-night effects are large.}
%..............................................................................
\InsertFigures{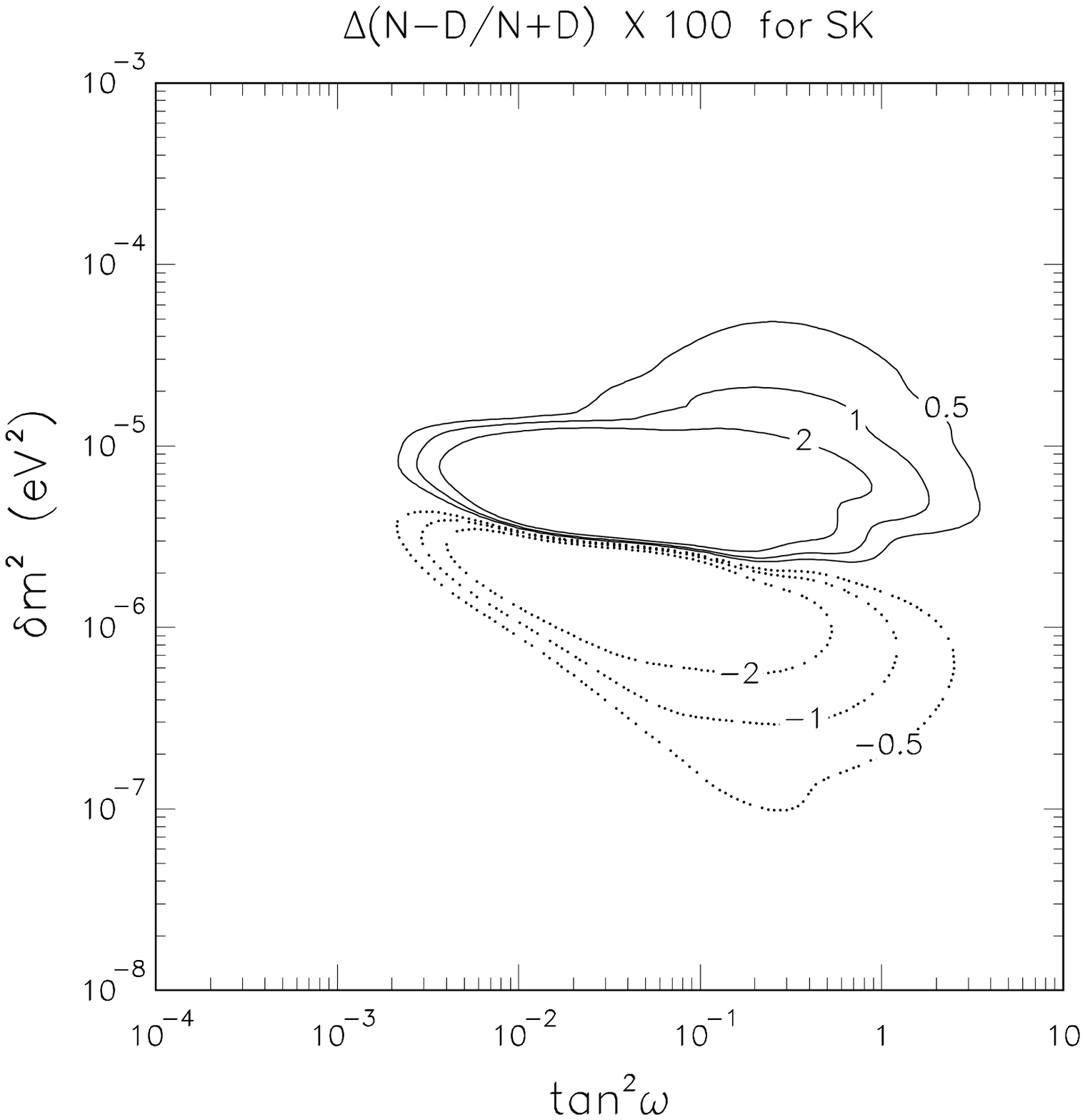}%
{Fig.~3. The difference $\Delta$ between the night-day asymmetry calculated in
the two ranges $[7.5,20]$ and $[5,7.5]$ MeV for the SK detector   ($\times
100$). Solid and dotted curves refer to $\Delta>0$ and $<0$, respectively. The
LOW and LMA solutions in Fig.~1 predict opposite signs for $\Delta$.}
%..............................................................................
\InsertFigures{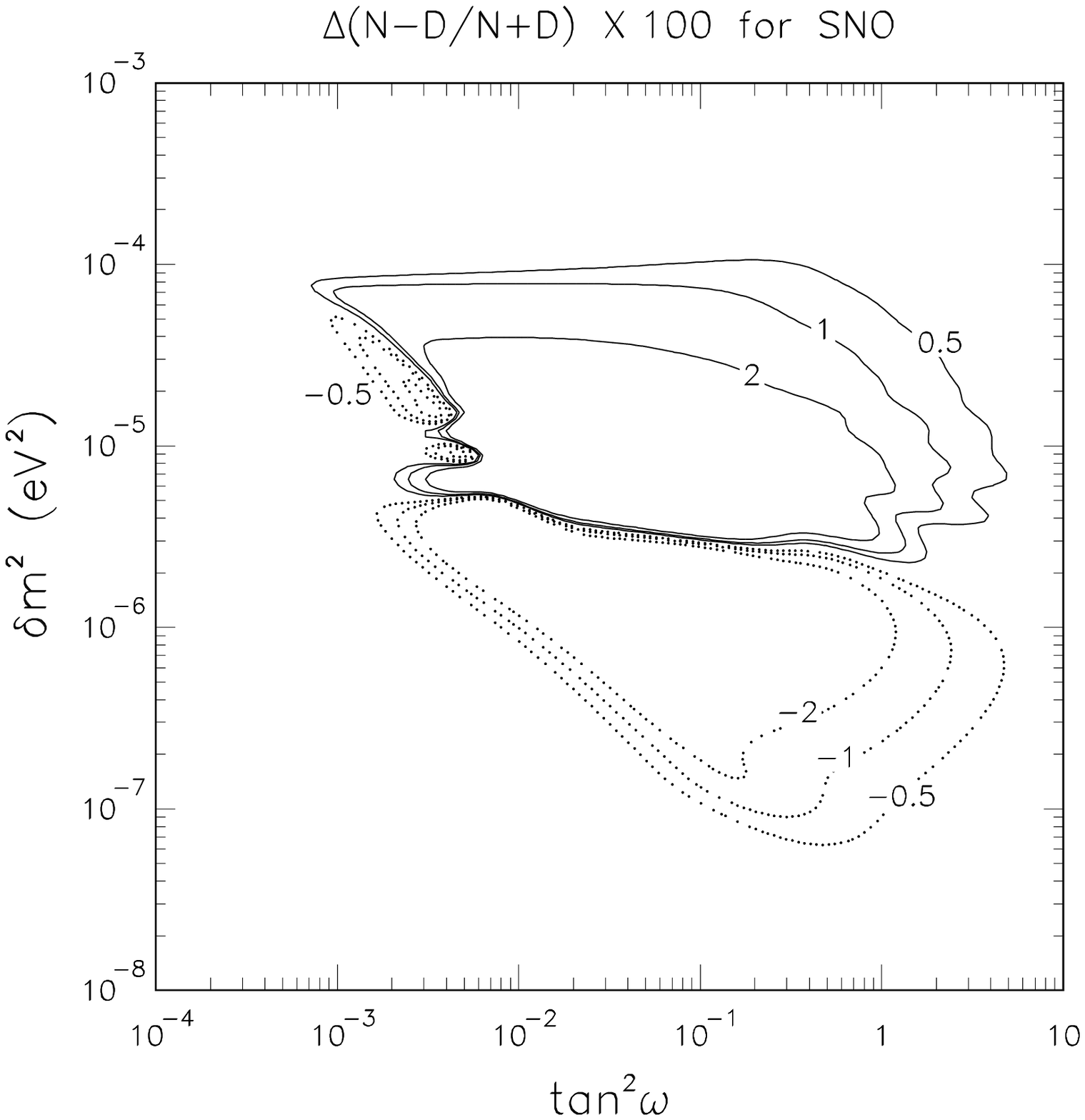}%
{\hfil Fig.~4. As in Fig.~3, but for the SNO detector.\hfil }

\eject
\end{document}